\documentstyle[emulateapj,epsf]{article}

\begin{document}

\title{ON THE MAXIMUM MASS OF DIFFERENTIALLY ROTATING NEUTRON STARS}

\author{Thomas W.~Baumgarte, Stuart L. Shapiro\altaffilmark{1},
	and Masaru Shibata\altaffilmark{2}}

\affil{Department of Physics, University of Illinois at
        Urbana-Champaign, Urbana, Il 61801}

\altaffiltext{1}{Department of Astronomy and NCSA, University of Illinois at
        Urbana-Champaign, Urbana, Il 61801}
\altaffiltext{2}{Department of Earth and Space Science,~Graduate School of
	Science,~Osaka University, Toyonaka, Osaka 560-0043, Japan}

\begin{abstract}
We construct relativistic equilibrium models of differentially
rotating neutron stars and show that they can support significantly
more mass than their nonrotating or uniformly rotating counterparts.
We dynamically evolve such ``hypermassive'' models in full general
relativity and show that there do exist configurations which are
dynamically stable against radial collapse and bar formation.  Our
results suggest that the remnant of binary neutron star coalescence may
be temporarily stabilized by differential rotation, leading to 
delayed collapse and a delayed gravitational wave burst.
\end{abstract}


\section{Introduction}

One of the most important characteristics of a neutron star is its
maximum allowed mass.  The maximum mass is crucial for distinguishing
between neutron stars and black holes in compact binaries and in
determining the outcome of many astrophysical processes, including
supernova collapse and the merger of binary neutron stars.

Observations of binary pulsars suggest that the individual stars in
such systems have masses very close to 1.4 $M_{\odot}$ (Thorsett et
al. 1993).  If mass-loss during the final binary coalescence can be
neglected, the remnant of such a merger will then have a {\em rest}
mass exceeding 3 $M_{\odot}$.  If this mass is larger than the maximum
allowed mass for neutron stars, then the merger will lead to prompt
collapse to a black hole on a dynamical timescale (ms).  If, however,
neutron stars can support such a high mass, at least temporarily, then
the merger may result in a high mass, quasi-equilibrium neutron star,
which only later may collapse to a black hole.  The two different
outcomes may have important consequences for gravitational wave
signals and, possibly, gamma ray burst models.

The maximum mass of a cold, nonrotating, spherical neutron star is
uniquely determined by the Tolman-Oppenheimer-Volkoff equations and
depends only on the cold equation of state.  For most recent equations
of state this maximum mass is in the range of 1.8 -- 2.3 $M_{\odot}$
(Akmal, Pandharipande and Ravenhall 1998), significantly smaller than
the mass expected for the remnant of a binary neutron star merger.

Thermal pressure and uniform rotation can provide additional
support, and may stabilize slightly more massive stars.  In this
paper, we point out that {\em differential} rotation can significantly
increase the maximum allowed mass of neutron stars and may temporarily
stabilize the remnant of binary neutron star mergers.  Recent fully
relativistic simulations of binary neutron star mergers (Shibata \&
Uryu 1999) show that merger remnants are indeed differentially
rotating (as suggested by several Newtonian simulations) and that they
may support masses much larger than the maximum allowed mass of
spherical stars.

In the case of a head-on collision from infinite separation, thermal
pressure alone may support the merged remnant of progenitors (Shapiro
1998).  Thermal pressure is likely to have a much smaller effect for
coalescence from the innermost stable circular orbit since shock
heating on impact is less pronounced, and will be dissipated by
neutrino emission in $\sim 10$ s.

Rotation can further increase the maximum allowed mass.  The maximum
mass of a {\em uniformly} rotating star is determined by the spin rate
at which the fluid at the equator moves on a geodesic and any further
speedup would lead to mass shedding.  This maximum mass can be
determined numerically and is found to be at most $\sim 20$\% larger
than the nonrotating value (e.g. Cook, Shapiro, \& Teukolsky, 1992,
hereafter CST; 1994, and references therein).  It is therefore
unlikely that uniform rotation could support the remnant of a binary
neutron star merger.  Rotating equilibrium configurations with rest
masses exceeding the maximum rest mass of nonrotating stars
constructed with the same equation of state are referred to as
``supramassive'' stars (CST).

The merger of a binary neutron star system, however, will not result
in a uniformly rotating object, especially since the neutron stars are
likely to be close to being irrotational before merger (Bildsten \&
Cutler 1992; Kochanek 1992).  The remnant is likely to be {\em
differentially} rotating (see Rasio \& Shapiro 1999 for discussion and
references; Shibata \& Uryu 1999).  The star's core may then rotate
faster than the envelope, and it is easy to imagine that such a star
could support a significantly larger mass than its uniformly rotating
counterpart (see also Ostriker, Bodenheimer \& Lynden-Bell, 1966,
where this effect was demonstrated for white dwarfs).  We refer
to equilibrium configurations with rest masses exceeding the maximum
rest mass of a uniformly rotating star as ``hypermassive'' stars.

In contrast to the maximum mass of the nonrotating and uniformly
rotating stars, the maximum mass of differentially rotating stars
cannot be uniquely defined, since the value will depend on the chosen
differential rotation law.  In principle, one might even construct an
extensive Keplerian disk around the equator of the star, possibly
increasing the mass of the star by large amounts.  Instead of
constructing such extreme configurations, we seek to determine whether
a reasonable degree of differential rotation can have a significant
effect on the maximum mass of neutron stars.

Here we adopt a polytropic equation of state and a simple rotation law
to explore the effects of differential rotation on the maximum mass.
We construct relativistic equilibrium models and find that even for
modest degrees of differential rotation the maximum mass increases
significantly, easily surpassing the likely remnant mass of a binary
neutron star merger.  We then evolve high-mass models dynamically in
full general relativity, and find that there do exist models which are
dynamically stable against both radial collapse and bar formation.
These are plausible candidates for binary neutron star remnants.

\section{Equilibrium}

We adopt a polytropic equation of state $P = K \rho_0^{1 + 1/n}$,
where $P$ is the pressure and $\rho_0$ the rest-mass density.  We take
the polytropic constant $K$ to be unity without loss of generality,
and choose the polytropic index $n = 1$~\footnote{Since $K^{n/2}$ has
units of length, all solutions scale according to $\bar M = K^{n/2}
M$, $\bar J = K^n J$, $\bar \Omega = K^{-n/2} \Omega$, etc, where the
barred quantities are physical quantities, and the unbarred quantities
are our dimensionless quantities corresponding to $K = 1$ (compare
CST).}.

Relativistic equilibrium models of rotating stars have been
constructed by several authors, including Butterworth \& Ipser (1975),
Friedman, Ipser \& Parker (1986), Komatsu, Eriguchi, \& Hachisu
(1989), CST, Bonazzola et al.~(1993), and Stergioulas \& Friedman
(1995).  A comparison between several different methods can be found
in Nozawa et al. (1998).  We use the numerical code developed by CST,
which is based on the formalism of Komatsu, Eriguchi, \& Hachisu
(1989).

Constructing differentially rotating neutron star models requires
choosing a rotation law $F(\Omega) = u^t u_{\phi}$, where $u^t$ and
$u_{\phi}$ are components of the four velocity $u^{\alpha}$, and
$\Omega$ is the angular velocity.  For simplicity we follow CST and
consider the rotation law $F(\Omega) = A^2 (\Omega_c - \Omega)$, where
$\Omega_c$ denotes the central angular velocity and where the
parameter $A$ has units of length.  Expressing $u^t$ and $u_{\phi}$ in
terms of $\Omega$ and metric potentials yields eq.~(42) in CST, or, in
the Newtonian limit,
$\Omega = \Omega_c / (1 + \hat A^{-2} \hat r^2 \sin^2 \theta)$.
%\begin{equation}
%\Omega = \frac{\hat A^2}{\hat A^2 + \hat r^2 \sin^2 \theta}
%\, \Omega_c.
%\end{equation}
%In eq.~(1) we have rescaled $A$ and $r$ in terms of the equatorial
Here we have rescaled $A$ and $r$ in terms of the equatorial
radius $R_e$: $\hat A = A/R_e$ and $\hat r = r/R_e$.  The parameter
$\hat A$ is a measure of the degree of differential rotation and
determines the length scale over which $\Omega$ changes.  Since
uniform rotation is recovered in the limit $\hat A \rightarrow
\infty$, it is convenient to parametrize sequences by $\hat A^{-1}$.

We construct axisymmetric differentially rotating models using a
modified version of the scheme adopted in CST.  Instead of fixing the
central density in the iteration scheme for each model, we fix the
maximum density.  This change allows us to construct higher mass
models in some cases, since the central density does not always
coincide with the maximum density, and hence may not specify a model
uniquely.  Given a value of $\hat A$, we construct a sequence of
models for each value of the maximum density by starting with a
static, spherically symmetric star and then decreasing the ratio of
the polar to equatorial radius, $R_{pe} = R_{p}/R_{e}$, in small
increments.  This sequence ends when we reach mass shedding (for large
values of $\hat A$), or when the code fails to converge (indicating
the termination of equilibrium solutions), or when $R_{pe} = 0$
(beyond which the star would become a toroid).

\vbox{ \vskip 0.00truecm
\centerline{\epsfxsize=7.5truecm
\epsfbox{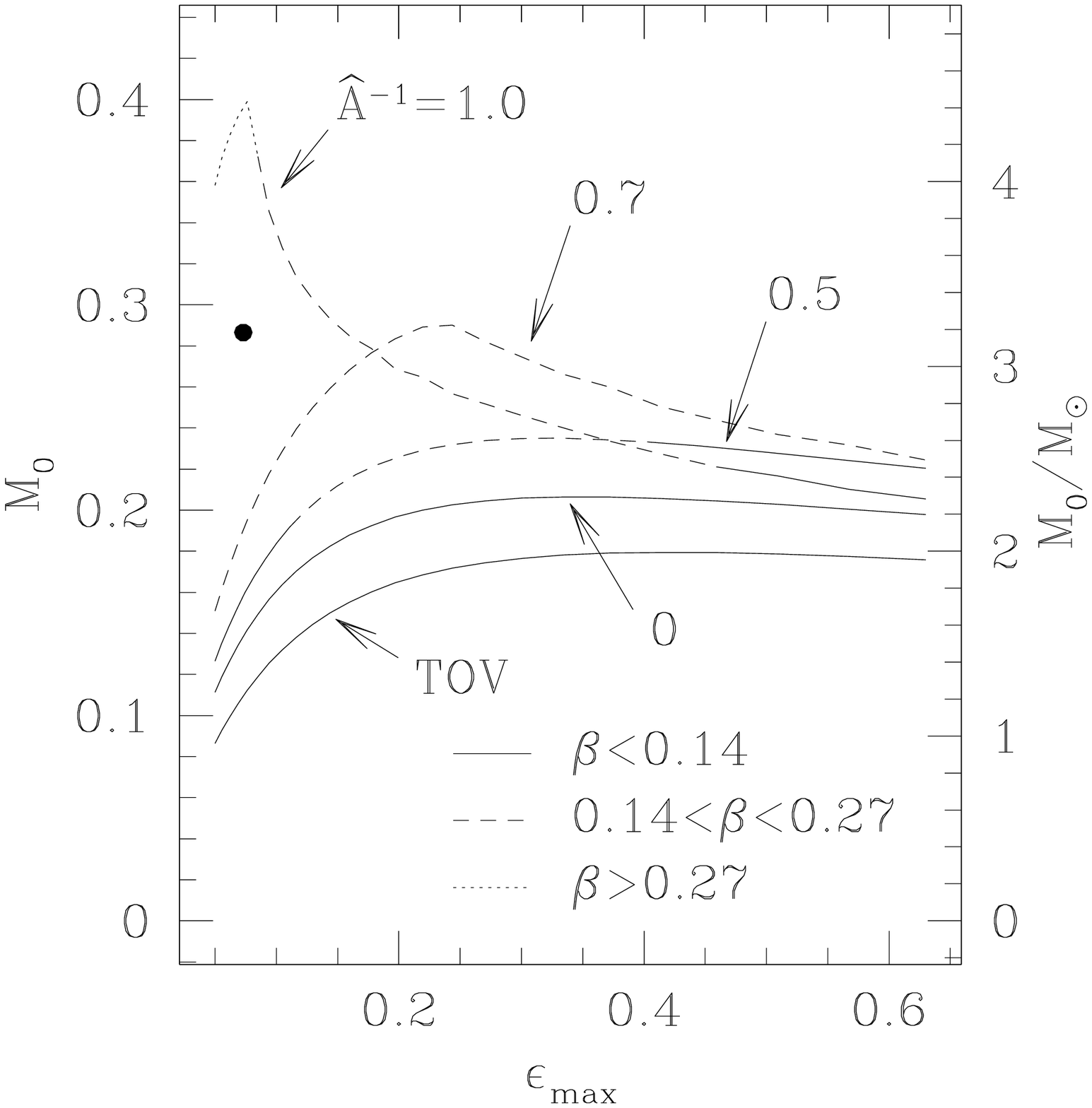}}
\vskip 0.00truecm 
\figcaption[]{ 
Maximum rest-mass configurations versus maximum mass-energy density
for differentially rotating $n=1$ sequences specified by $\hat
A^{-1}$.  Values of $\beta = T/|W|$ for the models are indicated. The
mass--density relation for static equilibrium stars (TOV) is shown for
comparison.  Masses in solar masses are calculated by assuming
that the maximum mass for nonrotating stars is 2 $M_{\odot}$.  Note
that even modest differential rotating may easily support $\gtrsim 3
M_{\odot}$, the expected mass of binary neutron star merger remnants.
We dynamically evolve the model marked with a dot, and show that it 
is dynamically stable (see Fig.~2).
} } \vskip 0.4truecm

In Fig.~1 we show the maximum-mass values in each sequence as a
function of the maximum value of the mass-energy density $\epsilon$
for different values of $\hat A$.  Even for modest differential
rotation, we can construct models with masses much higher than the
maximum mass for static and uniformly rotating stars.  Some of these
models exceed the Kerr limit $J/M^2 > 1$, where $J$ is the angular
momentum and $M$ the total mass energy of the star.

\section{Stability}

Nonrotating spherical stars are dynamically stable (unstable) against
radial modes if $\partial M/\partial \epsilon_c > 0$ ($\partial
M/\partial \epsilon_c < 0$), where $\epsilon_c$ is the central energy
density.  The same criterion can be applied to sequences of uniformly
rotating stars of constant $J$ to determine secular stability
(Friedman, Ipser, \& Sorkin 1988).  Exact criteria do not exist for
the dynamical stability of rotating stars; however, numerical
simulations of uniformly rotating models suggest that the onset of
dynamical stability is very close to the onset of secular instability
(Shibata, Baumgarte, \& Shapiro 1999a).

As an indication of the stability of our models against
nonaxisymmetric bar-mode formation, we have indicated values of the
ratio of their kinetic energy $T$ to potential energy $W$, $\beta
\equiv T/|W|$, in Fig.~1~\footnote{See~CST for relativistic
definitions of these quantities.}.  Newtonian stars develop bars on a
dynamical timescale when $\beta \gtrsim \beta_{\rm dyn} = 0.27$, while
they develop bars on a {\em secular} timescale for $\beta \gtrsim
\beta_{\rm sec} = 0.14$ via gravitational radiation or viscosity
(Chandrasekhar 1969, 1970; Houser, Centrella, \& Smith 1994).  For
relativistic stars, $\beta_{\rm sec}$ for gravitational wave-driven
bars is somewhat smaller than for Newtonian stars (Stergioulas \&
Friedman 1998), while $\beta_{\rm sec}$ for viscosity-driven bars is
slightly larger (Bonazzola, Frieben, \& Gourgoulhan 1996; Shapiro \&
Zane 1998).

\vbox{ \vskip 0.5truecm
\centerline{\epsfxsize=8.5truecm
\epsfbox{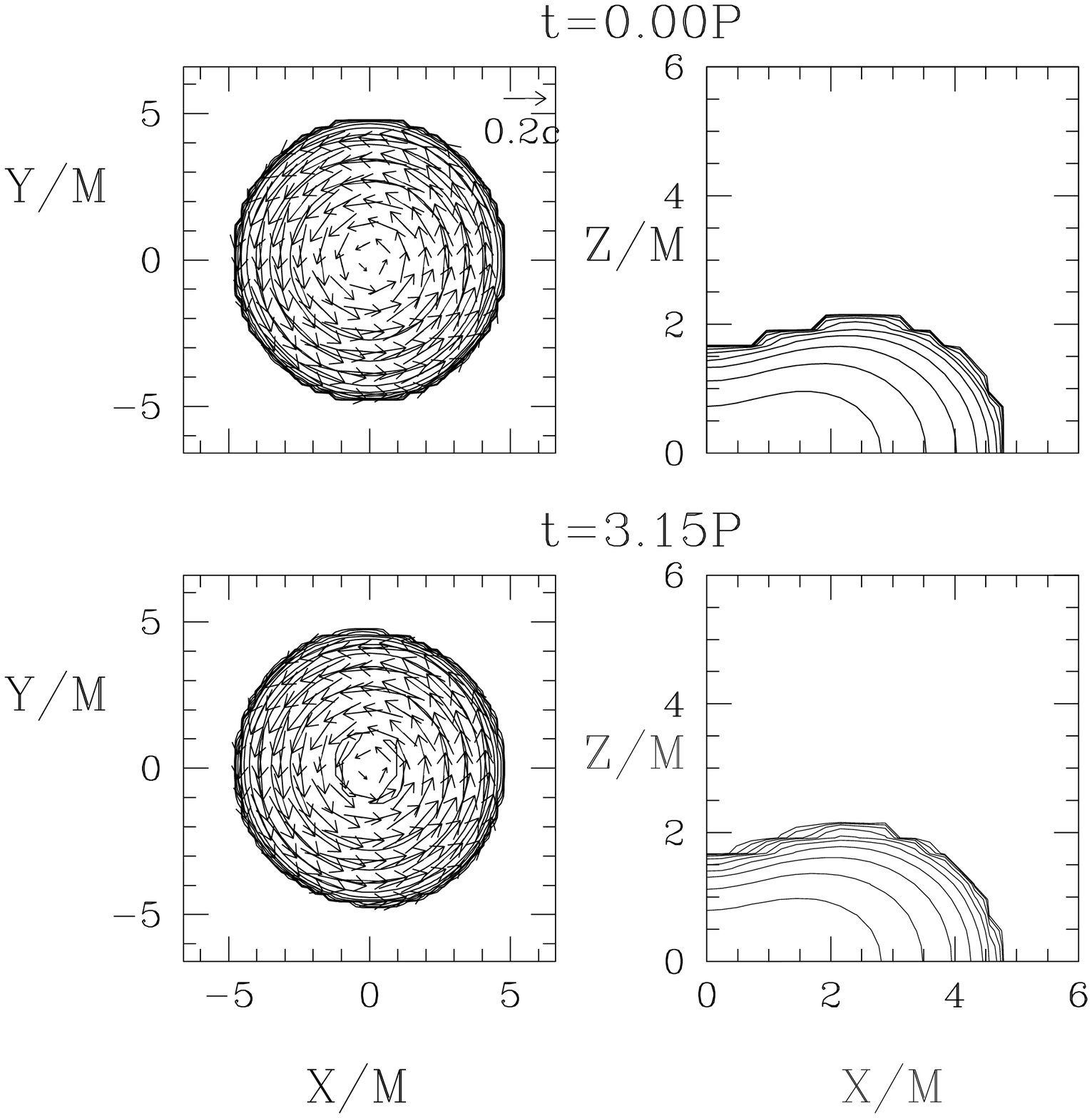}}
\vskip 0.00truecm \figcaption[]{ Snapshots of contours of the density
$\rho_0 u^0(-g)^{1/2}$ of the hypermassive star marked with a dot in
Fig.~1.  We show contours of the initial data and after about 3
central orbital periods, both in the equatorial
plane (left, including the velocity field $u^i/u^t$) and in a plane
containing the $z$-axis of rotation (right).  The star has a rest mass
of $M_0 = 0.29$, about $60\%$ larger than the maximum nonrotating rest
mass.  The simulation shows that this model is dynamically stable
against quasi-radial collapse and bar formation.  } }  \vskip
0.4truecm

To investigate the dynamical stability of our equilibrium models, we
insert them as initial data in a dynamical simulation and evolve them in
time.  We employ a fully relativistic code that solves Einstein's
equations coupled to hydrodynamics in three spatial dimensions plus
time (Shibata 1999).  

As a candidate for a dynamically stable star, we evolve the model with
$\hat A^{-1} = 1.0$, $\epsilon_{\rm max} = 0.073$ and $R_{pe} = 0.3$
(marked with a dot in Fig.~1).  This model has a rest mass about 60\%
higher than the maximum nonrotating rest mass, $\beta \sim 0.23$,
$R_e/M \sim 5$ and $J/M^2 \sim 1$, and is plotted in Fig.~2.  The
orbital period at the equator is about three times the orbital period
at the center.  We show contours at $t=0$ and after 3.15 orbital
periods at the center.  Clearly, this model is {\em dynamically}
stable against both quasi-radial collapse to a black hole and bar
formation, even when small perturbations are included initially.  This
demonstrates that differentially rotating stars can stably support
significantly higher masses than uniformly rotating stars for longer
than a dynamical timescale.  A more systematic study of the dynamical
stability of differentially rotating neutron stars will be presented
in a forthcoming paper (Shibata, Baumgarte, \& Shapiro 1999b).

Dynamically stable differentially rotating neutron stars are subject
to various {\em secular} instabilities.  The timescale for
gravitational-wave driven bar-mode formation can be estimated from
\begin{equation}
\tau_{\rm bar} \sim
	\left( \frac{M}{3 M_{\odot}} \right)^{-3}
	\left( \frac{R}{15 \mbox{km}} \right)^{4} 
	\left(\frac{\beta - \beta_{\rm sec}}{0.1}\right)^{-5} \mbox{s}
\end{equation}
(Friedman \& Schutz 1975, 1978), where the average radius $R$ and mass
$M$ are scaled to values appropriate for a binary merger remnant.  For
$\beta \sim 0.2$, this yields timescales of 10 s.  The final fate of
bar-unstable stars is not known, except for incompressible Newtonian
spheroids, where in the presence of gravitational radiation and
viscosity they evolve to Jacobian or Dedekind ellipsoids
(Chandrasekhar 1969; Miller 1974; Shapiro \& Teukolsky 1983; Lai \&
Shapiro, 1995).  Gravitational waves may also drive an $r$-mode
instability for arbitrarily small rotation rates (see, e.g., Lindblom,
Owen \& Morsink 1998).  For the hot remnants of binary neutron star
mergers, however, these modes may be suppressed by bulk viscosity.

Magnetic braking and viscosity will eventually bring the star into
uniform rotation.  When a hypermassive star is driven to uniform
rotation by viscosity or magnetic fields, it will undergo catastrophic
collapse and/or mass loss. The lifetime of a hypermassive star is
therefore set by these dissipative processes.  If $J/M^2 > 1$, angular
momentum must be dissipated either by radiation or mass loss before
the star can form a Kerr black hole (cf.~Baumgarte \& Shapiro 1998),
which may produce a massive, hot and thick disk around the newly formed
black hole.

A frozen-in magnetic field will be wound up by differential rotation,
which may create very strong toroidal fields.  This process will
generate Alfv\'en waves, which can redistribute and even carry off
angular momentum.  The timescale $\tau_B$ for this magnetic braking
mechanism is related to the Alfv\'en speed $v_A = B/(4\pi\rho)^{1/2}$
according to
\begin{equation}
\tau_B \sim \frac{R}{v_A} 
%	\sim \left( \frac{3M}{R} \right)^{1/2} \,\frac{1}{B}\\
	\sim 10^2
	\left(\frac{B}{10^{12} \mbox{G}}\right)^{-1}
	\left(\frac{R}{15 \mbox{km}}\right)^{-1/2}
	\left(\frac{M}{3M_{\odot}} \right)^{1/2} \mbox{s}.
\end{equation}
Here $B$ is the initial poloidal field along the gradient of $\Omega$.
Strong poloidal magnetic fields can increase the maximum allowed mass
of neutron stars (Bocquet et al. 1995) and contribute to the
dissipation of angular momentum by dipole radiation, but are subject
to a variety of MHD instabilities (e.g. Spruit 1999a, 1999b).

Since the fluid flow in differentially rotating equilibrium stars is
divergence-free, the viscous timescale $\tau_V$ is determined by
shear viscosity
\begin{equation}
\tau_V \sim \frac{\rho R^2}{4 \eta}
%	\frac{R^2 T^2}{4 \times 347 \rho^{5/4}} \sim
%	\frac{\pi^{5/4} 4^{1/4}}{3^{5/4} 347}\,\frac{R^{23/4}T^2}{M^{5/4}}
%	\\
	\sim 10^9
	\left(\frac{R}{15 \mbox{km}}\right)^{23/4}
	\left(\frac{T}{10^9 \mbox{K}}\right)^2
	\left(\frac{M}{3M_{\odot}} \right)^{-5/4}
	\mbox{s},\nonumber
\end{equation}
where $\eta = 347 \rho^{9/4} T^{-2}$ (cgs) (Cutler \& Lindblom 1987).
Molecular viscosity alone is likely to be less effective in bringing
the star into uniform rotation than magnetic braking.  Nascent neutron
stars may also be subject to convective instabilities (e.g.~Pons et
at.~1999), but the role of convection in rotating magnetic stars is
not well understood (cf.~Tassoul 1978).

For weak magnetic fields and high values of $\beta$, the neutron star
merger remnant is likely to develop a bar.  The accompanying
quasi-periodic gravitational wave signal may be observable by the new
generation of gravitational wave laser interferometers under
construction (Lai \& Shapiro 1995; Shibata, Baumgarte, \& Shapiro
1999b).

For strong magnetic fields and small values of $\beta$, magnetic
braking is likely to dominate the evolution of differentially
rotating neutron stars, and may alter the velocity profile within
minutes.  On this timescale, differential rotation will no
longer be able to support hypermassive stars formed in binary merger.
In the resulting delayed collapse, a brief secondary burst of
gravitational waves will be emitted.  The frequency of this secondary
burst may be quite high\footnote{The frequency of the fundamental
quasi-normal mode of a Schwarzschild black hole is $\omega \sim 0.37
M^{-1}$, which yields $f \sim 4$ kHz for $M = 3 M_{\odot}$; the
frequency of the axisymmetric mode is slightly higher for a Kerr black
hole (Leaver 1985).}, but since the angular momentum parameter $J/M^2$
may be close to unity, the amplitude could be large enough to be
observable by an advanced generation of gravitational wave detectors
(Stark \& Piran, 1985).  If the orbital parameters, including the
masses and radii of the stars, can be determined during the inspiral
and early merger phase, and if the time of the initial coalescence can
be inferred from the initial burst signal, then the measurement of this
delay in the final collapse may provide an estimate for the strength
of the wound-up magnetic field in the interior of the merged neutron
star.

\section{Discussion}

We find that the maximum mass of a differentially rotating star can be
significantly higher than that of nonrotating or uniformly rotating
stars, even for modest degrees of differential rotation.  As an
immediate consequence, it is possible that binary neutron star
coalescence does not lead to a prompt black hole formation, but that,
instead, a differentially rotating, hypermassive quasiequilibrium
neutron star is formed.  This has important consequences for the
gravitational wave signal from such an event and possibly for the
prospects of explaining gamma ray bursts by binary neutron star
mergers.

Pulsars are likely to be uniformly rotating, since magnetic braking
and viscosity will bring any initially differentially rotating stars
into uniform rotation.  The well established maximum masses of
uniformly rotating neutron stars are therefore relevant for old
neutron stars, including millisecond pulsars, while our much higher
maximum masses may be relevant for nascent neutron stars in a
transient phase in a supernova, following fallback, or in a merged
binary.

\acknowledgments

This work was supported by NSF Grants AST 96-18524 and PHY 99-02833
and NASA Grants NAG 5-7152 and NAG 5-8418 at the University of
Illinois at Urbana-Champaign.  Numerical computations were performed on
the FACOM VX/4R machine in the Data Processing Center of NAOJ, and at
the National Center for Supercomputing Applications at the University
of Illinois.  M.~Shibata gratefully acknowledges the kind hospitality
at the Department of Physics of the University of Illinois, and
support through a JSPS Fellowship for Research Abroad.

\end{document}